\documentclass[]{jpsj3}
\usepackage{txfonts}
\usepackage{cite}
\usepackage{color}

\title{Implementation of van der Waals Density Functional Approach 
to the Spin-Polarized System: Interaction Potential between Oxygen Molecules}

\author{\name{Masao \surname{OBATA}},$^1$ 
        \name{Makoto \surname{NAKAMURA}},$^1$
        \name{Ikutaro \surname{HAMADA}},$^2$
        and \name{Tatsuki \surname{ODA}}$^{1,3}$}

\inst{
$^1$ Graduate School of Natural Science and Technology, 
Kanazawa University,
Kakuma, Kanazawa 920-1192, Japan \\
$^2$ International Center for Materials Nanoarchitectonics (MANA), \\ 
National Institute for Materials Science (NIMS), Tsukuba 305-0044, Japan \\
$^3$ Institute of Science and Engineering, Kanazawa University, Kakuma, 
Kanazawa 920-1192, Japan} 

\abst{
We propose a practical approach to spin-polarized systems within the van der Waals density functional (vdW-DF).
The method was applied to a gas phase oxygen molecule and a parallel (H-type) pair of oxygen molecules.
It was found that vdW-DF improves the equilibrium distance and binding energy. In particular, one type of vdW-DF can describe such systems reasonably well.
The van der Waals interaction has been confirmed to have an energy comparable to the magnetic one, while emerging at a distance rather longer than the latter.
}

\kword{van der Waals, density functional theory, oxygen, intermolecular interaction, magnetic interaction}

\begin{document}
\maketitle
%%%%\section{Introduction}
%
The modeling of molecular systems has been a large research field for a long time. 
The van der Waals (vdW) force is an important issue in intermolecular 
interactions, as well as the covalent bonding for intramolecular interactions, 
electrostatic force, and magnetic force (atomic force, which originates from magnetic interaction) in spin-polarized systems. 
The van der Waals density functional (vdW-DF) of Dion {\it et al}.\cite{Dion2004} can be used to describe
the vdW forces within density functional theory (DFT), opening up a possibility
to describe a wider range of materials including molecular complexes,
layered materials, and adsorption systems (see e.g. Ref.~2 for review)
% layered materials, and adsorption systems (see e.g. ref.~\cite{Langreth2009} for a review)
%

%
While the number of applications of vdW-DF has increased,
much effort has been devoted to improve vdW-DF,
proposing new exchange and correlation functionals appropriate for vdW-DF.
Klime\v{s} {\it et al.}\cite{Klimes2010} optimized exchange functionals, 
which are compatible with the nonlocal correlation functional of vdW-DF.
Cooper\cite{Cooper2010} developed a functional form of the exchange functional
for vdW-DF.
%\textcolor{red}{
%Klime\v{s} {\it et al.}\cite{Klimes2010} and Cooper\cite{Cooper2010} propose exchange functionals,
%which are more appropriate than that used in the original vdW-DF.}
%
%\textcolor{red}{By optimizing the functional form with the density gradient,}
%Klime\v{s} {\it et al.}\cite{Klimes2010} and Cooper\cite{Cooper2010} proposes exchange functionals,
%which are compatible with the correlation functional of vdW-DF.
%
They provide better agreement than the original vdW-DF for the weakly 
interacting fragment in the benchmark S22 dataset.\cite{Jurecka2006}
Vydrov and Van Voorhis developed nonlocal correlation functionals
by making a variation in the model dielectric function within the plasmon-pole approximation
%\textcolor{red}{by making a variation in the dispersion model within the plasmon-pole approximation 
%on dielectric function}
and obtained more accurate results for the S22 dataset.\cite{Vydrov2009,Vydrov2010}
%
%\textcolor{red}{Simply, adjusting the screened exchange parameter,}
%
Lee {\it et al.}\cite{Lee2010} proposed the second version of vdW-DF (vdW-DF2)
that employs the refit Perdew-Wang exchange functional,\cite{Murray2009} which is free of spurious exchange binding and best mimics the Hartree-Fock exchange energy, and large-$N$ asymptote gradient correction to determine the vdW kernel, improving both equilibrium distance and interaction 
energy for extended as well as molecular systems.
%
%Lee {\it et al.}\cite{Lee2010} proposed a second version of vdW-DF (vdW-DF2),
%which shows improved results in both equilibrium distance and interaction 
%energy for extended as well as molecular systems. 
%
Hamada and Otani\cite{Hamada2010} proposed that, by combining Cooper's exchange (C09$_{x}$)
with vdW-DF2 (vdW-DF2$^{\text{C09}_{x}}$),
a more accurate description of graphite and graphene adsorption on transition metal surfaces is possible than in the case with the original vdW-DF.
They also showed that vdW-DF2$^{\text{C09}_{x}}$ predicts the equilibrium distance and adsorption energy of a water monomer on graphene,\cite{Hamada2012} which are comparable to those
obtained by the random phase approximation and quantum Monte-Carlo approaches.
More recently, vdW-DF based on the Baysian error estimation function with vdW correlation
was proposed and applied to a variety of systems.\cite{Wellendorff2012}
%
%For spin polarized systems no report has been published concerned with vdW-DF 
%theory and the spin polarized version of, for example, generalized gradient 
%approximation (GGA) has been popular so far.\cite{Perdew1992,Perdew1996}

Aside from the improvement of vdW-DF in describing accurate geometries of interaction energies
of materials, the extension of vdW-DF to spin-polarized (magnetic) systems may be important.
Thus far, the generalization of vdW-DF to spin-polarized systems has never been reported,
except that by Vydrov and Van Voorhis.\cite{Vydrov2009}
Once this extension is established, we will be able to study intriguing magnetic systems,
such as molecular magnets and hybrid metal-organic interfaces, with vdW-DF.
Although the direct magnetic interaction, which plays an important role in describing
magnetic properties of the system, extends within a short range, obtaining accurate
intermolecular configurations or adsorption geometries with vdW forces is crucial,
as they determine the (interfacial) electronic structure.\cite{Morikawa2004, Toyoda2010, Hamada2012}
As a first approach to spin-polarized systems, 
the use of the oxygen molecule and related compounds may be appropriate,
because of the spin triplet ground state and relatively 
simple ($2s$, $2p$) electronic structure of the molecule.
In gas and liquid phases, the role of vdW forces between molecules 
has been frequently discussed so far.\cite{Ewing1975,Aquilanti1999} 
Theoretically, intermolecular interactions have been treated
by using a quantum chemistry method\cite{Bussery1993,Hernandez1995,Nozawa2002}
and molecular dynamics techniques.\cite{Oda2002,Doreman1984}
In the former, the complete-active-space self-consistent field (CASSCF) level of theory is
most reliable, but its application is limited to small systems.
The latter is based on the parametrized model of Lennard-Jones potential or 
on DFT, in which electron potential is described in a nonempirical fashion.
The latter level of approximation results in an overestimation of intermolecular interaction energy
and an unlikely compacting structure for pairing molecules.
In particular, because the energy scale of magnetic interaction is comparable 
to the transition temperature among gas, liquid, and solid phases 
at normal pressures,\cite{Santoro2001,Oda2004} a more reliable description 
of the vdW interaction is desirable to study condensed oxygen systems.
In solid phases at high pressures, it is expected that the vdW and magnetic interactions will 
compete with the chemical bonding energy. 
Moreover, the oxygen molecule may be frequently discussed in terms 
of the vdW intermolecular interaction with many kinds of materials. 

%%% In the theory of quantum chemistry, many approaches have been reported 
%%% for O2-O2 inter-molecular potential. XXXXXXXXXXXXXXXXX

In this letter, we propose a practical approach to spin-polarized systems within vdW-DF.
We examined our approach by applying it to the oxygen-oxygen interacting 
pair  (O$_{2}$)$_{2}$ with the parallel (H-type) configuration, which is frequently found in a 
local structure in condensed oxygen.
The intermolecular potential energy curve was investigated for several vdW-DFs,
and it was found that one of them predicts equilibrium intermolecular distance and interaction energy,
in good agreement with experiments.
The roles of vdW and magnetic interactions are also discussed.
%

%%%\section{Method} 

The exchange-correlation (XC) functional in vdW-DF is written as
\begin{equation}
E_{\rm xc}             = E_{\rm x}+E_{\rm c}^{\rm loc}+E_{\rm c}^{\rm nl},  
\label{eq1}
\end{equation}
where $E_{\rm x}$ and $E_{\rm c}^{\rm loc}$ are the exchange functional
and short range local contribution to the electron correlation, respectively.
The last term on the right-hand-side is the nonlocal correlation given by
\begin{equation}
E_{\rm c}^{\rm nl}[n] =  \frac{1}{2}\iint d{\bf r}d{\bf r}^{\prime} 
                                   n({\bf r}) \phi({\bf r}, {\bf r}^{\prime}) n({\bf r}^{\prime}),
\label{eq2} 
\end{equation}
where $\phi$ is the vdW kernel, which satisfies the following properties:
$E_{\rm c}^{\rm nl}$ is strictly zero for a homogeneous electron gas,
and the interaction between two fragments has the correct $R^{-6}$ dependence
for a large separation,\cite{Dion2004} where $R$ is the distance between the fragments.
$E_{\rm c}^{\rm nl}$ is a functional of the electron density $n({\bf r})$ and its gradient $\vert \nabla n \vert$.
This property allows us to directly extend the term to spin polarized systems 
by setting $n  =  n_{\uparrow}+n_{\downarrow}$, where $n_{\uparrow}$ and 
$n_{\downarrow}$ are the spin-up and spin-down components of the spin density, respectively.
However, for spin polarized systems, the other contribution 
to $E_{\rm xc}$ in eq.~({\ref{eq1}}) has been determined in terms of both electron and spin densities. 
As for $E_{\rm x}$, the revised Perdew-Burk Ernzerhof (revPBE)\cite{Zhang1998} flavor of the exchange functional within the generalized gradient approximation (GGA) has been used in the original vdW-DF.\cite{Dion2004}
As in previous work on vdW-DF,
we will test some choices for $E_{\rm x}$ in the next section. 
$E_{\rm c}^{\rm loc}$ in eq.~(\ref{eq1}) for spin-polarized systems may be written as
\begin{eqnarray}
E_{\rm c}^{\rm loc} & = & E_{\rm c}^{\rm LSDA}[n_{\uparrow}, n_{\downarrow}],
\label{eq4} 
\end{eqnarray}
where $E_{\rm c}^{\rm LDA}$ is the (short-range) correlation energy within the local spin density (LSDA) approximation.
There is an ambiguity in the choice of $E_{\rm c}^{\rm loc}$.
In the original vdW-DF, the local density approximation (LDA) was used 
to avoid possible double counting of the contribution of $\vert \nabla n \vert$ contained in $E_{\rm c}^{\rm nl}$.
%
%However, in the case of spin-polarized systems, the term involves a contribution from density gradient and spin polarization.
%
To take the gradient correction (GC) with spin polarization into account, we propose a local part of the correlation functional as follows:
\begin{equation}
E_{\rm c}^{\rm loc}= E_{\rm c}^{\rm LSDA}[n_{\uparrow},n_{\downarrow}] + \Delta E_{\rm c}[n,\zeta],
\label{eq6}
\end{equation}
where
\begin{eqnarray}
\Delta E_{\rm c}[n,\zeta]&=&E_{\rm c}^{\rm PBE}[n_{\uparrow}, n_{\downarrow}] - E_{\rm c}^{\rm PBE}[n/2, n/2]  \nonumber \\
                                 &=&\int\! d{\bf r}\ n\ [H(r_{s},\zeta ,t)-H(r_{s},0,t)], 
\label{eq7}
\end{eqnarray}
$H$, $r_{\rm s}$, $\zeta$, and $t$ are the gradient contribution, Seitz radius ($n=3/4\pi r_{s}^{3}$), 
spin polarization ratio ($\zeta=(n_{\uparrow}-n_{\downarrow})/n$), and 
dimensionless density gradient proportional to $|\nabla n|$, respectively.\cite{Perdew1996}
$\Delta E_{\rm c}$ depends on both $\zeta$ and $|\nabla n|$,
and vanishes when there is no spin polarization.
Thus, the present functional reduces to the original one in the absence of spin polarization.
We note that the importance of gradient correction in the local part of the correlation was
also pointed out.\cite{Wellendorff2011}

The self-consistent treatment has been employed for the Kohn-Sham equation\cite{Kohn1965} $\left( T + V_{\rm KS} \right) \Psi_{i}  =  \varepsilon_{i} \Psi_{i}$, 
where $\Psi_{i}$ and $\varepsilon_i$ are the $i$-th one-electron wave function 
and eigenvalue for the Hamiltonian, which contains the kinetic energy operator $T$ and 
the potential $V_{\rm KS}$.
The latter is evaluated in a self-consistent way from a set of 
wave functions.
The correlation potential for the proposed approach (eqs.~\ref{eq6} and \ref{eq7}) is written as
\begin{eqnarray}
V_{\rm c}^{\sigma} &\! =\! & v_{c}^{\sigma } + v_{c}^{nl} + \frac{\delta \Delta E_{\rm c}}{\delta n_{\sigma}},
\label{eq10}
\end{eqnarray}
%
%The correlation potential for the original vdW-DF (eq.~\ref{eq4}) is written as
%\begin{eqnarray}
%V_{\rm c}^{\sigma} & = & v_{c}^{\sigma}+v_{c}^{nl},
%\label{eq9} 
%\end{eqnarray}
where $\sigma$ indicates the spin-up~($\uparrow$) or spin-down~($\downarrow$) states,
and $v_{c}^{\rm nl}=\delta E_{\rm c}^{\rm nl}/\delta n$.
%\begin{eqnarray}
%V_{\rm c}^{\sigma} &\! =\! & v_{c}^{\sigma } + v_{c}^{nl} + \frac{\delta \Delta E_{\rm c}}{\delta n_{\sigma}}.
%\label{eq10}
%\end{eqnarray}
%
The contribution of the nonlocal part is practically calculated 
using the algorithm developed by Rom\'{a}n-P\'{e}rez and Soler.\cite{Roman-Perez2009} 
This algorithm considerably accelerates and simplifies the calculation of
$E_{\rm c}^{\rm nl}$ and $v_{c}^{\rm nl}$, 
in which the computational cost scales $N\log N$ ($N$ is the number of real space grid points),
while the calculation of $E_{\rm c}^{\rm nl}$ in the original formula
needs an operation proportional to $N^2$.\cite{Dion2004} 
In our implementation, the modified approach by Wu and Gygi\cite{Wu2012} was used
with the following parameters: $q_{\rm c}=8, N_q=31,$ and $m_{\rm c}=12$. 
As for the $q$-points, we used a linear mesh for a few initial values,
and for the remaining points, a log-mesh was used as in the original work.\cite{Roman-Perez2009}
This treatment enables us to use $q=0$ on a $q$-point grid.
%This enables us to define the grid on the zero of $q$. 
%

%
Associated with a plane wave basis in the computational program,\cite{Laasonen1993} 
a cubic box with a dimension of 10.6~\AA\ was used for oxygen cluster systems. 
In the construction of ultrasoft-pseudopotentials\cite{Vanderbilt1990} with the atomic code,
we used the PBE functional 
and neglected the contribution of $E_{\rm c}^{\rm nl}$ in the XC.
The nonlocal correlation was explicitly included in the molecular calculations
with the plane wave code.
This treatment has been found to suffer from few effect
in physical quantities obtained  from the total energy difference
in clusters and solid systems.\cite{Klimes2011,Hamada2012}
Using the self-consistently determined wave functions, atomic forces can be calculated.
Plane wave cutoffs of 40 and 350 Ry were used for wave functions and electron 
densities, respectively.
We used PBE, the original vdW-DF,\cite{Dion2004} 
vdW-DF$^{\text{C09}_{x}}$,\cite{Cooper2010} vdW-DF2$^{\text{C09}_{x}}$, \cite{Hamada2010}
and those with GC (denoted as, for e.g., vdW-DF-GC) as XC functionals .
We confirmed that the program code developed in the present work
reproduces the binding energy curves of graphite\cite{Hamada2010}
and the interaction energy curve of the benzene-water pair in the S22 dataset.\cite{Wu2012}

%%%\section{Result}

%-----------------------O2-----------------------------
%%%\subsection{Oxygen Molecule}

\begin{figure}
\begin{center}
\includegraphics[width=8.0cm]{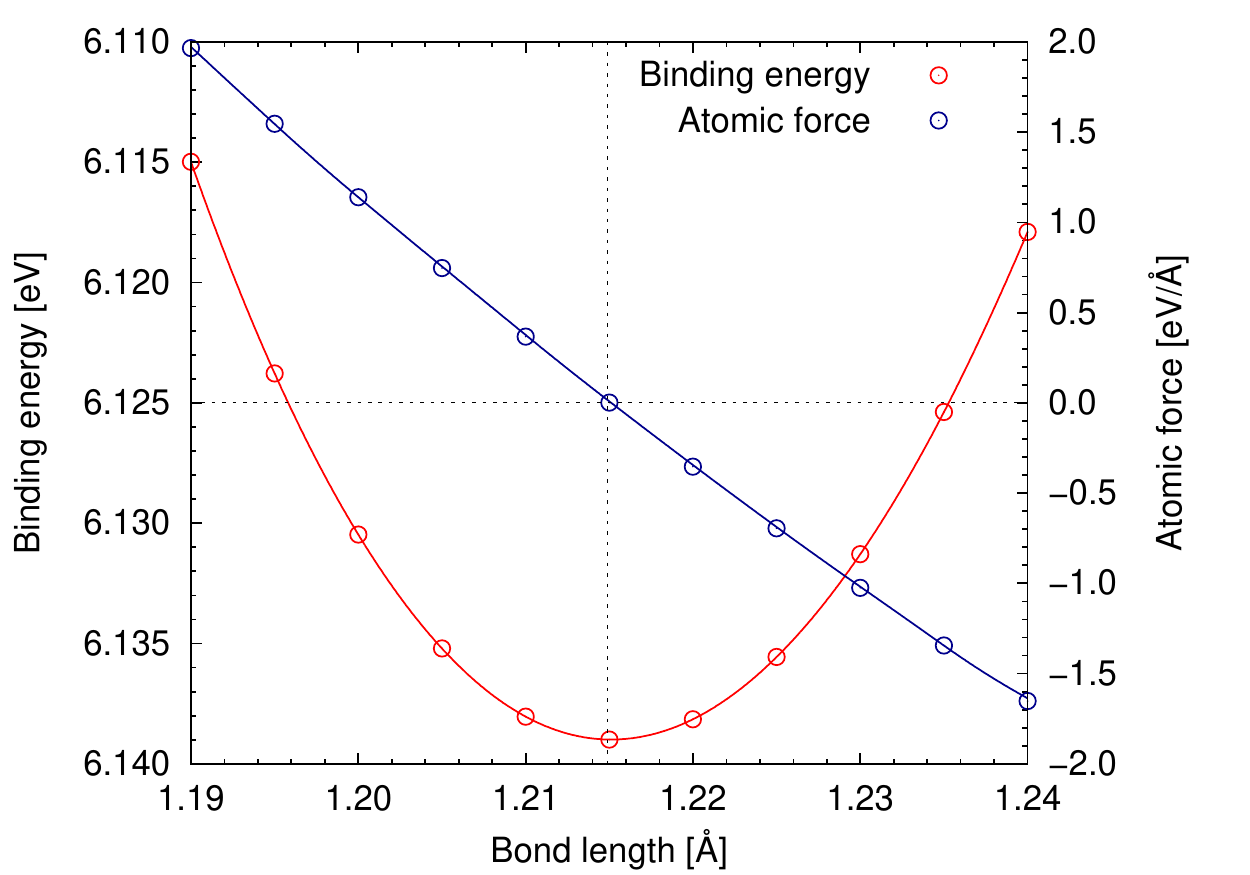}
\caption{
(Color) Binding energy and atomic force of oxygen molecule
as functions of bond length.
vdW-DF2$^{\text{C09}_{x}}$-GC was used for the calculations.
}
\label{fig1}
\end{center}
\end{figure}

Table~\ref{table1} shows the calculated properties of an oxygen molecule with several functionals.
The calculated equilibrium bond length ($b$), binding energy ($E_{\rm b}^{\rm mol}$) ,
and vibration frequency ($\omega$) are comparable to the experimental values. 
Interestingly, the data obtained with vdW-DFs shows that introduction of GC,
defined by eq.~(\ref{eq7}), lowers the binding energy by 0.23~eV.
$E_{\rm b}^{\rm mol}$ obtained using vdW-DF-GC is much smaller than that obtained using PBE,
and in better agreement with the experimental value, but $b$ is overestimated.
vdW-DFs with the C09 exchange tend to decrease $b$, leading to better agreement with the experimental value than
PBE and vdW-DFs with the revPBE exchange, but overestimate $E_{\rm b}^{\rm mol}$ significantly.
Nevertheless, $E_{\rm b}^{\rm mol}$'s obtained using PBE and vdW-DFs with the C09 exchange are in reasonably good agreement.
Figure~\ref{fig1} shows binding energy and atomic force 
as functions of bond length for the oxygen molecule calculated using vdW-DF2$^{\text{C09}_{x}}$-GC.
The equilibrium bond length is estimated to be 1.215~\AA\ from the energy minimum. 
This differs by less than $1 \times 10^{-3}$~\AA\ from the bond length determined from the atomic force.
The vdW-DF approach does not drastically change the equilibrium bond length, 
binding energy, vibration frequency, or the energy gap
between the highest occupied molecular orbital (HOMO) and 
the lowest unoccupied molecular orbital (LUMO) levels, compared with  results of GGA (PBE).
Our data is in line with the fact that the XC functional of vdW-DFs has a predictive power
for small molecules similar to GGA.\cite{Thonhauser2007}

\begin{table}
\caption{Properties of oxygen molecule: equilibrium bond lengths~($b$), 
binding energies~($E_{\rm b}^{\rm mol}$), 
vibrational frequencies~($\omega$), and  
HOMO-LUMO energy gaps~($\Delta \varepsilon_{\rm H-L}$).}
\label{table1}
\begin{center}
\begin{tabular}{l|cccc}
\hline \hline
energy functional               & $b$\ (\AA) & $E_{\rm b}^{\rm mol}$\ (eV) & 
        $\omega$(THz)         & $\Delta \varepsilon_{\rm H-L}$\ (eV)        \\
\hline
PBE                           & 1.221 &  6.03 & 44.7 &  2.38 \\
vdW-DF                        & 1.233 &  5.50 & 45.7 &  2.26 \\
vdW-DF-GC                     & 1.232 &  5.27 & 45.6 &  2.37 \\            
vdW-DF$^{\text{C09}_{x}}$     & 1.214 &  6.49 & 48.4 &  2.22 \\
vdW-DF$^{\text{C09}_{x}}$-GC  & 1.214 &  6.26 & 48.3 &  2.33 \\
vdW-DF2$^{\text{C09}_{x}}$    & 1.215 &  6.37 & 48.3 &  2.22  \\
vdW-DF2$^{\text{C09}_{x}}$-GC & 1.215 &  6.14 & 48.3 &  2.33   \\
Exp.                          & 1.207$^a$ & 5.12$^a$ & 47.39$^a$ &   \\
\hline
\end{tabular}
$^a$ Ref.~\cite{Huber1979}
\end{center}
\end{table}

%% Several properties in O2 (equilibrium bond length, binding energy, 
%% vibrational frequency, HOMO-LUMO energy gap)

%%%-----------------------O2-O2-----------------------------
%%%\subsection{Pair of oxygen molecules}

\begin{figure}
\begin{center}
\includegraphics[width=8.0cm]{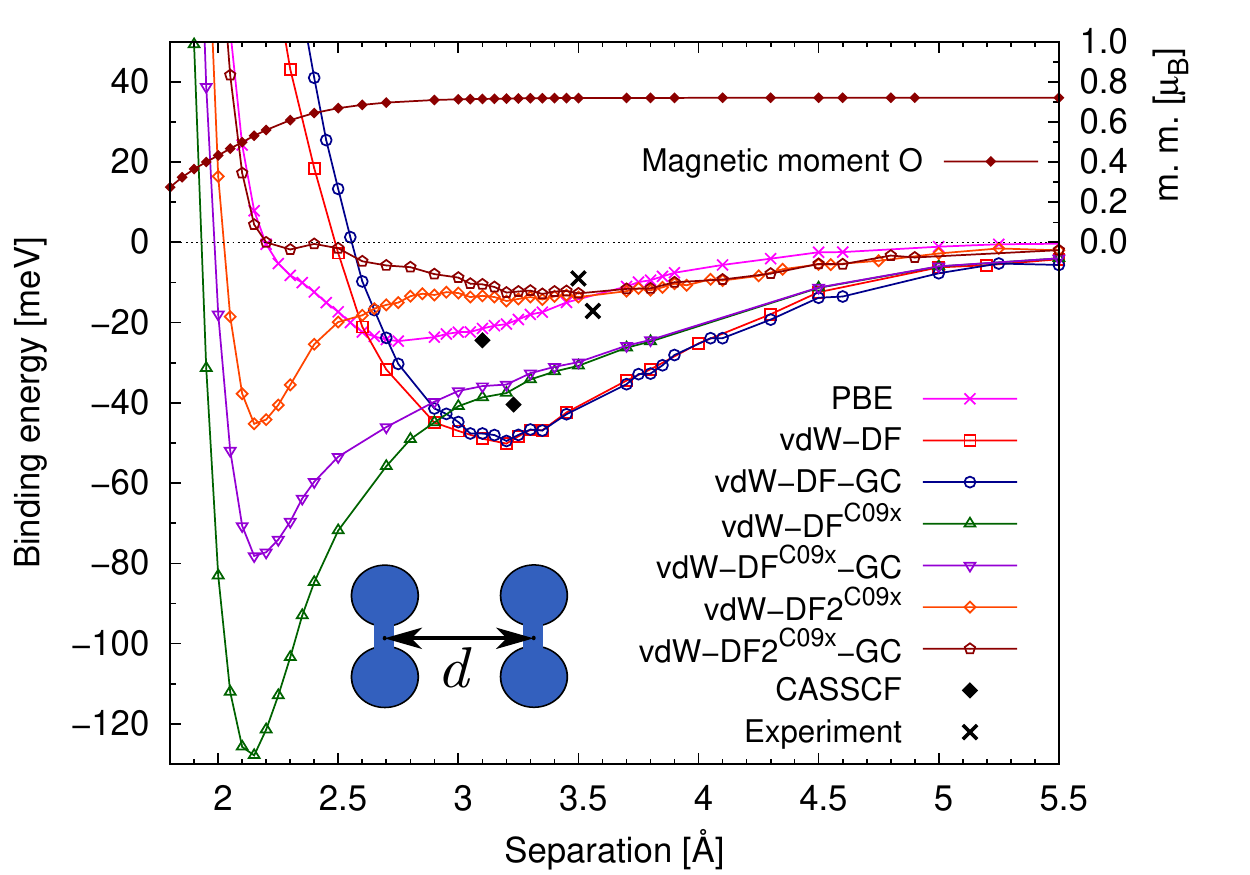}
\caption{
(Color) Intermolecular interaction energy as a function of the separation between oxygen molecules
at the antiferromagnetic parallel-molecule configuration (see the inset for the configuration),
together with the results from experiment and quantum chemistry calculation.
The atomic magnetic moment (m.m.) is also reported in the upper part of the panel.
}
\label{fig2}
\end{center}
\end{figure}

The intermolecular interaction potential was evaluated in the 
parallel (H-type) molecular configuration with several functionals. 
In the calculation, the bond length in each molecule was fixed to 
its respective ground-state value listed in Table~\ref{table1}. 
Figure \ref{fig2} shows the binding energy as a function of the distance between centers of molecules ($d$).
The reference total energy was set to that of the molecular pair at $d$ larger than 6.4 \AA.
This energy differs by only 0.5 meV from twice the total energy 
minimum of single molecule. 
The properties of the binding energy curve are shown in Table~\ref{table2}. 
The equilibrium distance and binding energy were determined
by fitting the binding energy curve around the minimum to a third-order polynomial.

\begin{figure}
\begin{center}
\includegraphics[width=8.0cm]{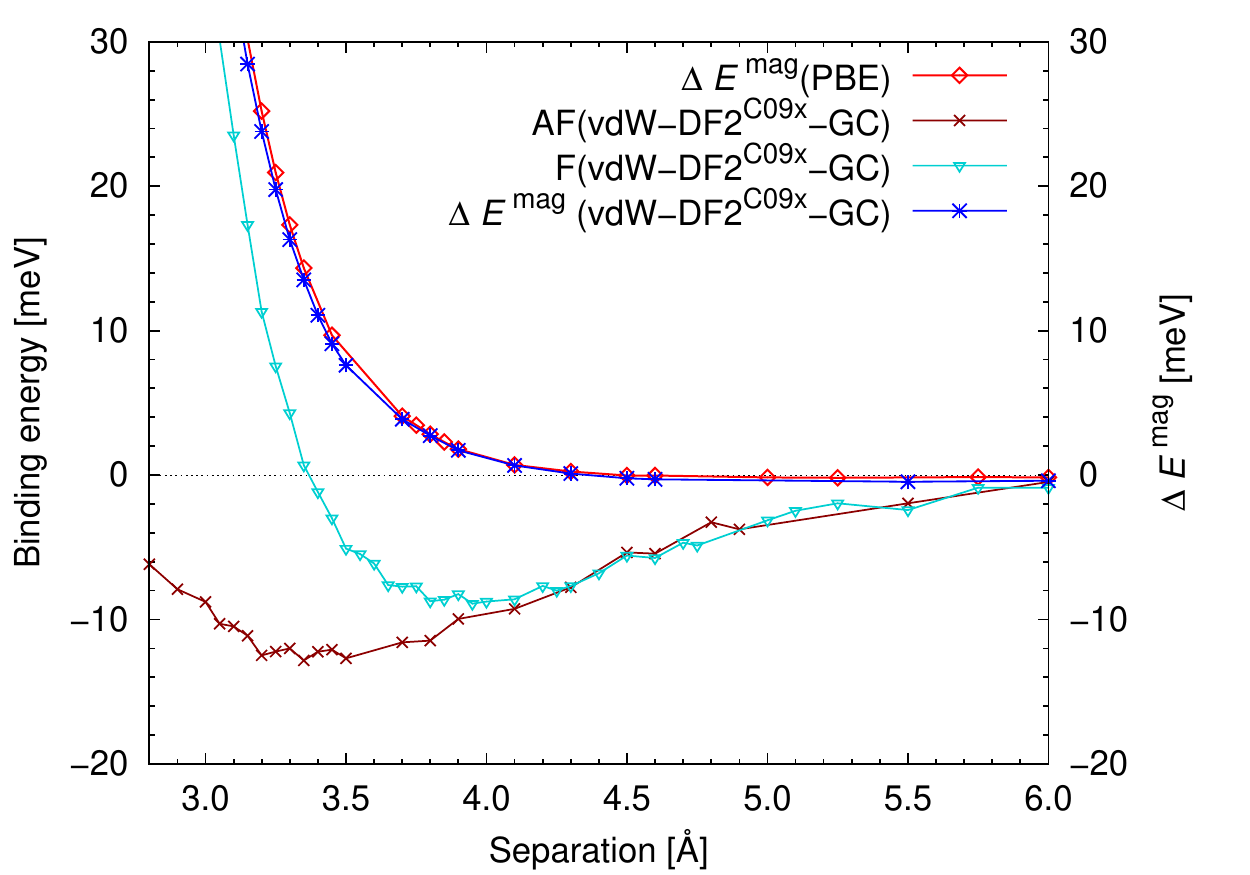}
\caption{
(Color) Magnetic interaction energies with respect to the intermolecular distance 
and the intermolecular interaction energies of the antiferro~(AF)- and ferro~(F)-magnetic 
coupling oxygen molecules are also shown.}
\label{fig3}
\end{center}
\end{figure}

The potential curves from the original vdW-DF  show a lower energy
than those with GC and a strong repulsive feature at distances smaller than 3.5~\AA. 
This repulsive interaction with GC makes the equilibrium distance larger,  
especially in vdW-DF2$^{\text{C09}_{x}}$. 
The repulsive force is attributed to the intermolecular magnetic interaction.
At approximately 2.15~\AA\, such a repulsive nature becomes weak in accordance with the 
decrease in magnetic moment (see upper part of Fig.~\ref{fig2}, which is generally 
common among functionals), resulting in an energy minimum or dip in the potential 
curves.
Note that the repulsive potential, which appears at less than 2~\AA, is mainly of nonmagnetic origin. 
With the original vdW-DF with GC, the minimum is observed at 3.20~\AA,
with vdW-DF$^{\text{C09}_{x}}$-GC at 2.16~\AA, 
and with vdW-DF2$^{\text{C09}_{x}}$-GC at 3.44~\AA. 
The first and last functionals almost reproduce the distance obtained in 
the previous work, which employs the quantum chemistry method or 
experimental measurement: 3.1~\AA, 3.23~\AA~(CASSCF),\cite{Hernandez1995,Nozawa2002}
and 3.5~\AA, 3.56~\AA~(experiment).{\cite{Ewing1975,Aquilanti1999}}
Concerning the binding energy of the O$_{2}$ pair ($E_{\rm b}^{\rm O_{4}}$), 
the value calculated using vdW-DF2$^{\text{C09}_{x}}$-GC (12.9~meV){\cite{Comment}} is in good agreement
with experimental values; 9~meV,\cite{Ewing1975} 17.1~meV.\cite{Aquilanti1999} 
The value obtained using vdW-DF-GC (48.0~meV) is similar to the result by CASSCF (40.4~meV), 
as indicated by the feature of equilibrium distance.
These agreements with the more elaborate method or the recent experiment 
imply that vdW-DF-GC and vdW-DF2$^{\text{C09}_{x}}$-GC 
predict a desirable potential for describing the electronic structures of larger systems,
%%such as (O$_{2}$)$_{8}$ cluster and solid oxygen.
such as (O$_{2}$)$_{4}$ cluster and solid oxygen.

%%Quantum chemistry: complete active space self-consistent 
%%field method(CASSCF)
%%Experimental measurement: Quantum interface scatterring(QIS)

The magnetic interaction energy between molecules may be given
by the energy difference $\Delta E^{\rm mag}$ between ferro~(F)- and 
antiferro~(AF)-magnetic coupling pairs: 
$\Delta E^{\rm mag} = E^{\rm F} - E^{\rm AF}$, where $E^{\rm F}$ and $E^{\rm AF}$ are the total energies for the F and AF states, respectively.
Figure~\ref{fig3} shows $\Delta E^{\rm mag}$ as a function of the separation $d$
obtained using vdW-DF2$^{\text{C09}_{x}}$.
$\Delta E^{\rm mag}$ obtained using PBE is also shown for comparison.
The functional dependence of the magnetic interaction within vdW-DF was found to be small (not shown). 
$\Delta E^{\rm mag}$ emerges at 4.1~\AA\ and increases by 24~meV at 3.2~\AA.
Such a magnetic energy scale is comparable to the binding energy of the O$_{2}$ pair,
implying that  the energy minimum in the potential curve 
(vdW-DF-GC or vdW-DF2$^{{\rm C09}_x}$-GC) 
is determined not only by the balance between the Pauli repulsion and the vdW attraction,
but also by an intermolecular magnetic interaction.  
%
%In PBE energy curve, as shown in Fig.~\ref{fig2}, 
%the separation distance of minimum energy changes shorter owing to 
%the lack of vdW attraction. 
%%
%This means that the stability due to AF magnetic coupling appears 
%around $d=2.8 {\rm \AA}$ while those of vdW do around 3.2$-$3.5~\AA.   
%
Note that the $\Delta E^{\rm mag}$'s obtained
using vdW-DF2$^{{\rm C09}_x}$ and PBE are almost identical.
Because PBE lacks a long-range vdW attraction,
the result suggests that our present approach (vdW-DF with GC) describes
the short-range magnetic interaction as accurate as PBE.

%%% 2-2. Behavior at the potential tail: ${\rm R}^{-6}$ dependence

The tail of the interaction energy curve is characterized by the function $-C_{6}/d^{6}$. 
The coefficient $C_{6}$ obtained by fitting the calculated data (Table~\ref{table2})
varies from 47 to 121~eV\AA$^{6}$ in vdW-DFs.
The values in the data are dispersive,
but decrease in order of publication year of the functionals. 
These values are comparable to or larger than
those of Ar (37~eV \AA$^{6}$) and N$_{2}$ (47~eV \AA$^{6}$) dimers.\cite{Dion2004, Hult1991}
%%% This coincidence can interestingly 
%%% call a similarity on the gas-liquid transition temperature at normal pressure.
%
We note that $C_{6}$ coefficients should be identical when the same nonlocal correlation functional
is used, because, in the vdW asymptote, $C_{6}$ does not depend on exchange or short-range
correlation energy, but on charge density and $q_0$ function.\cite{Dion2004,Vydrov2009}
Thus, the difference between the $C_{6}$ coefficients (for e.g., difference in those obtained with vdW-DF and vdW-DF$^{{\rm C09}_x}$) may be attributed to the use of a small simulation cell,
and the differences (within 27 eV${\rm \AA}^{6}$) should be regarded as numerical errors.
As for the Lennard-Jones potential, the hard-core 
diameter $\sigma_{0}$ of the intermolecular distance where the potential energy vanishes is worth investigating.
As indicated in Table~\ref{table2},  $\sigma_{0}$'s with vdW-DFs are equal to or less than 
2.6~\AA.
The largest value is in agreement with those 
from theoretical calculations (2.7~\AA, 3.0~\AA)\cite{Nozawa2002,Bussery1993} 
while the smaller value at approximately 2~\AA\ offers a picture of smaller particles 
in the gas phase.
    
%% 2-3. Comparison with results of the configuration interaction 
%% in quantum chemistry calculation

%% HOMO-LUMO energy splitting

%%For discussion and simplicity, in this form, we consider the electronic 
%%levels for LUMO and HOMO only. In Fig. x, the energy difference 
%%between HOMO and LUMO levels is reported. This shows that

\begin{table}
\caption{Features of the intermolecular potential energy curve for vdW functionals  (vdW-DF, vdW-DF$^{{\rm C09}_x}$, vdW-DF2$^{{\rm C09}_x}$) and 
those with gradient correction (-GC): 
equilibrium distances~($d$), binding energies~($E_{\rm b}^{{\rm O}_{4}}$), $C_{6}$ coefficients, 
and hard-core diameters~($\sigma_{0}$). 
For comparison, the data of PBE, CASSCF and  
experimental measurement are also listed.}
\label{table2}
\begin{center}
\begin{tabular}{l|cccc}
\hline \hline
energy functional           & $d$ (\AA)            & $E_{\rm b}^{{\rm O}_{4}}$ (meV) & 
$C_{6}$ (eV\AA$^{6}$)  & $\sigma_{0}$(\AA)      \\
\hline
PBE                        &  2.77     &  24.6     &   22  &  2.20     \\
vdW-DF                     &  3.18     &  50.3     &  108  &  2.49     \\
vdW-DF-GC                  &  3.20     &  48.0     &  121  &  2.56     \\            
vdW-DF$^{{\rm C09}_x}$     &  2.13     & 128.3     &   95  &  1.93     \\
vdW-DF$^{{\rm C09}_x}$-GC  &  2.16     &  78.5     &   94  &  1.98     \\
vdW-DF2$^{{\rm C09}_x}$    &  2.16     &  45.5     &   47  &  2.02     \\
vdW-DF2$^{{\rm C09}_x}$-GC &  3.44     &  12.9     &   48  &  2.20     \\
CASSCF                     &  3.1$^a$  &  24.4$^a$ &       &           \\
                           &  3.23$^b$ &  40.4$^b$ &       &  2.7$^b$  \\
Exp.                       &  3.5$^c$  &   9$^c$   &       &           \\
                           &  3.56$^d$ &  17.1$^d$ &       &           \\
\hline
\end{tabular}
$^a$ Ref.~\cite{Hernandez1995}, $^b$ Ref.~\cite{Nozawa2002}, $^c$ Ref.~\cite{Ewing1975}, $^d$ Ref.~\cite{Aquilanti1999}
\end{center}
\end{table}

%%%\section{Discussion}

%%%\section{Summary}

In summary, we have proposed a practical approach to spin-polarized systems within vdW-DF
and applied it to the spin-triplet oxygen molecules.
For the O$_{2}$ pair of H-type configuration, the vdW interaction 
has been discussed in comparison with the magnetic interaction between molecules.
It was found that the combination of vdW with the magnetic 
interaction is responsible for describing the magnetic molecular complexes. 
It was also found that one type of vdW-DF (vdW-DF2$^{{\rm C09}_x}$-GC) 
offers a fairly satisfactory result in terms of both equilibrium distance and binding energy 
for the antiferromagnetic coupling oxygen molecules, when compared with the 
experimental data. 
The properties of vdW-DF-GC also 
found to be similar to those of the quantum chemistry method,
which provides a larger equilibrium distance than GGA.
These agreements with vdW-DF will initiate further study of systems comprising oxygen molecules. 
The application to extended systems is under investigation.
%%%, showing a stabilized favor in the solid oxygen. 

The results in this work indicated a potential of the vdW-DF approach for spin-polarized 
systems as well as nonmagnetic ones.
Our finding that vdW-DF can reproduce 
the experimental result between interacting magnetic molecules opens a way 
to the study of magnetic vdW complex systems without models parameterized for particular systems.
Our device for spin-polarized systems must be replaced 
with a spin polarization version of $E_{\rm c}^{\rm nl}$
as in the correlation functional of GGA.

%\begin{acknowledgment}

The computation in this work was performed using the facilities of 
the Supercomputer Center, Institute for Solid State Physics, 
University of Tokyo and the facilities of the Research Center for 
Computational Science, Okazaki, Japan. 
This work was partly supported by Grants-in-Aid for Scientific 
Research from JSPS/MEXT (Grant Nos. 22104012 and 22340106),
the Strategic Programs for Innovative Research (SPIRE), MEXT, Japan,
the Computational Materials Science Initiative (CMSI), Japan,
and by the World Premier International Research Center Initiative (WPI)
for Materials Nanoarchitectonics, MEXT, Japan

%%For environments for acknowledgement(s) 
%%are available: \verb|acknowledgment|, \verb|acknowledgments|, 
%%               \verb|acknowledgement|, and \verb|acknowledgements|.

%%\end{acknowledgment}

%%%\appendix
%%%\section{}

%%Use the \verb|\appendix| command if you need an appendix(es). 
%%The \verb|\section| command should follow even though there 
%%is no title for the appendix (see above in the source of this file).

%%For authors of Invited Review Papers, the \verb|profile| command is 
%%prepared for the author(s)' profile.  A simple example is shown below.
%%\begin{verbatim}
%%\profile{Taro Butsuri}{was born in Tokyo, Japan in 1965. ...}
%%\end{verbatim}

\end{document}